\documentclass[aps,pra,twocolumn,longbibliography]{revtex4-1}

\usepackage{graphicx}
\usepackage{bm,bbm}
\usepackage{xcolor} 
\usepackage{hyperref}
\usepackage{tcolorbox}
\definecolor{mycolor}{rgb}{0.122, 0.435, 0.698}
\tcbuselibrary{theorems}
\newtcbtheorem{Definitions}{Definition}%
{colback=mycolor!5,colframe=mycolor,fonttitle=\bfseries,
left=4pt,
right=4pt,
top=5pt,
bottom=5pt}{defi}
\newtcbtheorem{Observations}{Observation}%
{colback=green!5,colframe=green!50!blue,fonttitle=\bfseries,
left=4pt,
right=4pt,
top=5pt,
bottom=5pt,
title={#2}
}{lemma}
\newtcbtheorem{Results}{Result}%
{colback=purple!5,colframe=purple!50!blue,fonttitle=\bfseries,
left=4pt,
right=4pt,
top=5pt,
bottom=5pt
}{lemma}
\begin{document}

\title{The Schmidt number of a quantum state cannot always be device-independently certified}

\author{Flavien Hirsch}
\affiliation{Institute for Quantum Optics and Quantum Information - IQOQI Vienna, Austrian Academy of Sciences, Boltzmanngasse 3, 1090 Vienna, Austria}%
\author{Marcus Huber}
\affiliation{Institute for Quantum Optics and Quantum Information - IQOQI Vienna, Austrian Academy of Sciences, Boltzmanngasse 3, 1090 Vienna, Austria}%


\begin{abstract}
One of the great challenges of quantum foundations and quantum information theory is the characterisation of the relationship between entanglement and the violation of Bell inequalities. It is well known that in specific scenarios these two can behave differently, from local hidden-variable models for entangled quantum states in restricted Bell scenarios, to maximal violations of Bell inequalities not concurring with maximal entanglement. In this paper we put forward a simple proof that there exist quantum states, whose entanglement content, as measured by the Schmidt number, cannot be device-independently certified for all possible sequential measurements on any number of copies. While the bigger question: \textit{can the presence of entanglement always be device-independently certified?} remains open, we provide proof that quantifying entanglement device-independently is not always possible, even beyond the standard Bell scenario.
\end{abstract}

\maketitle
One of the cornerstones and pivotal results in quantum information is the 1989 paper by Reinhard Werner \cite{Werner}, constructing a local hidden-variable model of an entangled quantum state for all projective measurements. Since then, lots of effort has been invested into the characterisation of entanglement \cite{JensReview,HorodeckiReview,OtfriedReview,NicoReview} and Bell-inequality violations (aka 'non-locality') \cite{BrunnerReview}. One of the biggest open questions concerns the potential for any entangled state to violate some form of a Bell-inequality, or in other words, whether all entanglement can be device-independently certified.

For non-sequential measurements on single copies (aka 'the standard Bell scenario'), the answer is known to be negative \cite{Barrett02}, i.e. there exist entangled quantum states for which all possible positive operator valued measures (POVMs) on single copies can be explained by a local hidden-variable model. Processing multiple copies or allowing for sequences of measurements, however, opens a plethora of further options \cite{Peres96,Masanes06,Junge10,Palazuelos12,Miguel11,PopescuHNl,FlaHNL,Gallego14}. The simplest example being entanglement distillation, since any distillable state can, by definition, be distilled (close) to a pure entangled state, which all violate the CHSH inequality (bipartite case) \cite{Gisin91}, or another Bell inequality \cite{Popescu92,Mariami17} in the multipartite case. Therefore, the existence of bound entanglement, i.e. entanglement that cannot be distilled, was a contender for answering this question negatively, i.e. if such undistillable states could not violate Bell inequalities, then also many copies wouldn't have helped. However, contrary to what Peres had conjectured, there are examples of states positive under partial transposition (PPT), and thus bound entangled \cite{Horodecki_1998}, that can be semi- \cite{Moroder14} or fully device-independently \cite{Vertesi14} certified. It has also been shown recently that there exist PPT states which cannot violate any Bell inequality in the standard Bell scenario, but nevertheless can violate CHSH after suitable local filtering \cite{Tendick19}.
\begin{figure*}[t]
  \centering
 \includegraphics[width=0.9\textwidth]{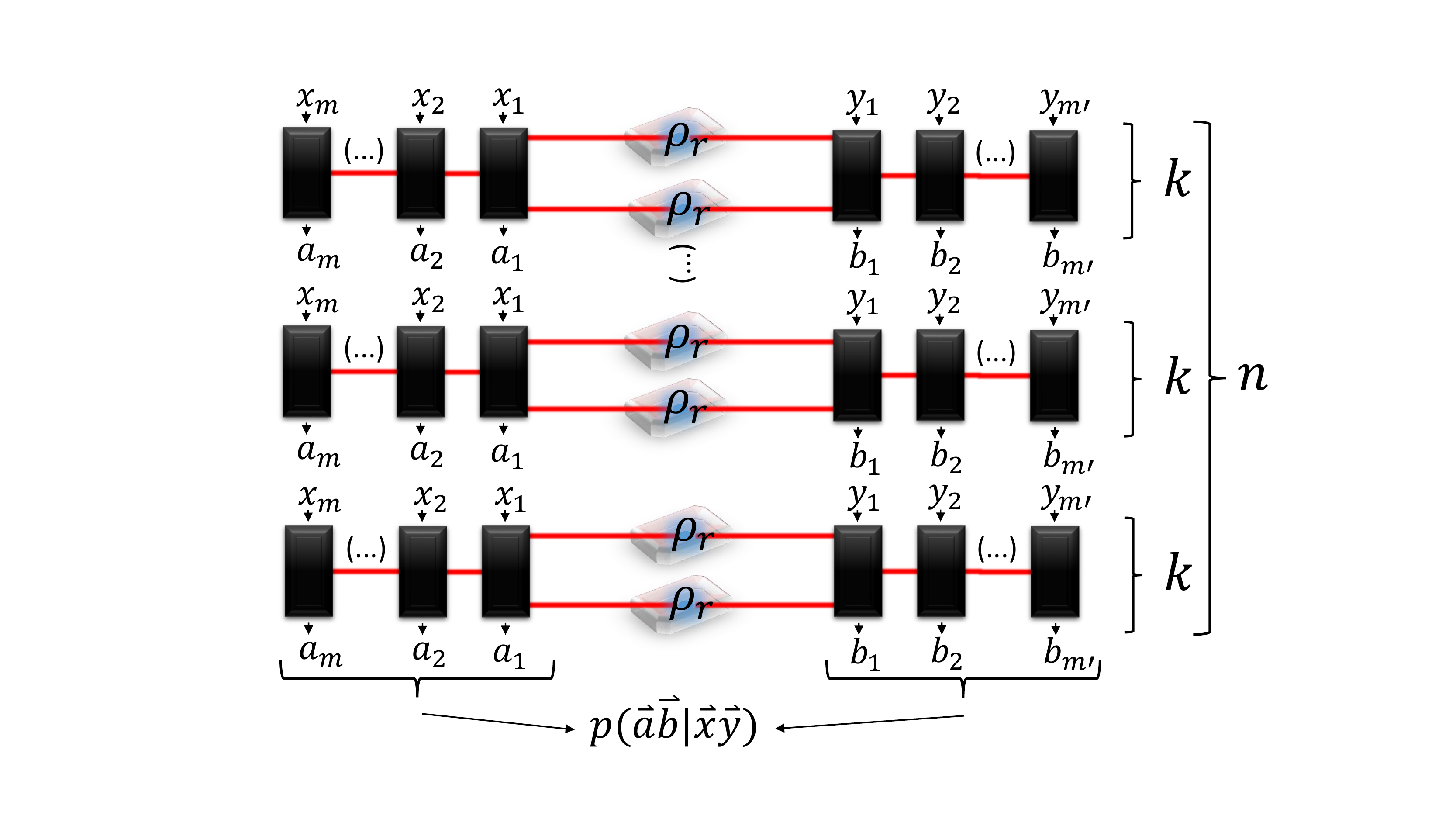}  
  \caption{We consider the paradigm in which we want to device-independently quantify the entanglement of a source capable of producing a quantum state $\rho_r$ with Schmidt number $r(\rho)$. In principle, the device can be used $n$ times and $k$ instances of the state are measured by POVMs that we also allow to be done sequentially. No restrictions are put on the number of inputs or outputs, yielding the joint probability distribution $p(\vec{a}\vec{b}|\vec{x}\vec{y})=p(a_1a_2\cdots a_m b_1b_2\cdots b_m|x_1x_2\cdots x_m y_1y_2\cdots y_m)$ for $n\rightarrow\infty$.}
  \label{fig:Schematic}
\end{figure*}

While we don't know the answer to the question on the presence of entanglement, there are multiple examples in which maximal entanglement does not coincide with maximal Bell violations beyond qubits \cite{Acin2012,Salavrakos2017}. This suggests, that the quantification of entanglement may not be possible device-independently. An example where it is possible, is the so called self-testing scenario \cite{Mayers98}. Here, up to isometries, exact states and with it their entanglement content can be perfectly certified from Bell-inequality violations (assuming quantum mechanics). While all pure states can be self tested  \cite{Coladangelo217}, these techniques usually suffer from a poor resistance to experimental noise \cite{Supi2019selftesting}. An easier variant is the use of Bell inequalities for certifying lower bounds on entanglement measures \cite{Moroder2013}, which has predominantly been used on a single copy. But imperfect certification or self-testing does not preclude the existence of better schemes that work on multiple copies of the quantum state to certify the entanglement device-independently and perfectly. To answer this question negatively one would need an example in which all possible (sequential) measurements on arbitrarily many copies can be explained by quantum states with less entanglement than the states actual entanglement content. As a figure of merit, we consider the Schmidt number \cite{Terhal} which was previously proven to be device-independently certifiable in some cases in \cite{Brunner08}. 

In this short letter we leverage a recent result on the existence of states whose partial transpose are also valid quantum states, with a different Schmidt number \cite{PPT} to provide examples, in which correctly certifying the Schmidt number is impossible in any conceivable device-independent scenario. To start, we need to introduce the notion of a bounded Schmidt number model:

\begin{Definitions*}{$k$-dimensional hidden quantum model}{ }
A probability distribution, that can be obtained by measurements on a state of Schmidt number $k$
\begin{align}
    p(ab|xy)=\text{Tr}(M_{a|x}\otimes M_{b|y}\rho_k)\,,
\end{align}
is defined to have a \emph{$k$-dimensional hidden quantum model} ($k$-HQ model). 
\end{Definitions*}

Here, $M_{a|x}$ denotes a POVM element with setting $x$ and outcome $a$ and a state of Schmidt number $r(\rho)$ means $\rho_k$ can be decomposed into pure states of Schmidt rank at most $k$, i.e. the Schmidt number is defined as $r(\rho):=\text{inf}_{[\{p_i,|\psi_i\rangle\}s.t. \sum_ip_i|\psi_i\rangle=\rho]} \text{max}_i\text{rank}(\text{Tr}_B[|\psi_i\rangle\langle\psi_i|])$ \cite{NicoReview} (with $p_i\geq0$).

In other words, it is possible that the obtained probability distribution $p(ab|xy)$ has a quantum mechanical origin using entangled states of Schmidt number $k$. To device-independently certify a Schmidt number $k'>k$ of a quantum state $\rho_{exp}$, it is thus necessary to find measurements $\tilde{M}_{a|x}$ and $\tilde{M}_{b|y}$, such that the resulting probability distribution $p_{exp}$ does not admit a \emph{$k$-dimensional hidden quantum model}, ie.
\begin{align}
   \text{Tr}(\tilde{M}_{a|x}\otimes \tilde{M}_{b|y}\rho_{exp})=p_{exp}(ab|xy)\neq\text{Tr}(M_{a|x}\otimes M_{b|y}\rho_k)\,.
\end{align}
\begin{Observations*}{Hidden quantum models and partial transpose}{}
 The existence of a \emph{$k$-HQ} model for a probability distribution coming from a state $\rho$ and measurements $\tilde{M}_{a|x}, \tilde{M}_{b|y}$, implies a \emph{$k$-HQ} model for the partially transposed state $\rho^{T_B}$ and measurements $\tilde{M}_{a|x}$, $\tilde{M}^T_{b|y}$.
\end{Observations*}
\textbf{Proof:} Let's write the probability distribution admitting a \emph{$k$-HQ} model for some state $\rho_{exp}$
\begin{align} \label{proof}
    p(ab|xy)=\text{Tr}(\tilde{M}_{a|x}\otimes \tilde{M}_{b|y}\rho_{exp})=\text{Tr}(M_{a|x}\otimes M_{b|y}\rho_k)\nonumber\\=\text{Tr}(\tilde{M}_{a|x}\otimes \tilde{M}^T_{b|y}\rho_{exp}^{T_B})=\text{Tr}(M_{a|x}\otimes M_{b|y}\rho_k)\,,
\end{align}
where used the fact $\text{Tr}(XY^{T_B})=\text{Tr}(X^{T_B}Y)$, i.e. the self-duality of the partial transpose map.
Now let's assume a state positive under partial transposition (PPT) $\rho_{PPT}^{T_B}\geq0$ and $r(\rho_{PPT})=m$ and $r(\rho_{PPT}^{T_B})=n<m$. For $\rho_{PPT}$ it follows that for all possible measurements $\tilde{M}_{a|x}$ and $\tilde{M}_{b|y}$ there exists a \emph{$n$-HQ} model. Indeed,  for all $\tilde{M}_{a|x}$ and $\tilde{M}_{b|y}$
\begin{align}
    \text{Tr}(\tilde{M}_{a|x}\otimes \tilde{M}_{b|y}\rho_{PPT})=\text{Tr}(M_{a|x}\otimes M_{b|y}\rho_n)\,,
\end{align}
where $M_{a|x}=\tilde{M}_{a|x}$, $M_{b|y}=\tilde{M}^T_{b|y}$ and $\rho_n=\rho_{PPT}^{T_B}$. In other words, any possible measurement procedure on the state of Schmidt number $m>n$ will produce probability distributions that can equivalently be obtained from measurements on a state of Schmidt number $n$. 

Now considering the following family of states, first considered in \cite{Ishizaka_2004} (for an even $d$), 
\begin{align}
    \rho(d)=\frac{(\mathbbm{1}_4-\omega_2)\otimes(\mathbbm{1}_{d^2/4}-\omega_{d/2})+(d/2+1)\omega_2\otimes\omega_{d/2}}{3d^2/4+d/2-2}\,,
\end{align}
where $\omega_d$ is the projector on the maximally entangled state in dimension $d$, i.e. $\omega_d=|\phi^+\rangle\langle\phi^+|$  with $|\phi^+\rangle=\frac{1}{\sqrt{d}}\sum_{i=0}^{d-1}|ii\rangle$. In Ref.~\cite{PPT}, it was proven that this family of states
has a Schmidt number linear in the local dimension, i.e. $r(\rho(d))\geq \lceil \frac{d}{4} \rceil$. More remarkable in our context, the partial transpose only has a Schmidt number of $r(\rho^{T_B}(d))\leq4$, independent of $d$. 
This implies:
\begin{Results*}{Impossibility of device-independent Schmidt number certification}{}
 For any Schmidt number greater than 4, there exists a corresponding state  whose Schmidt number cannot be device-independently certified for any set of sequential measurements on any number of copies.
\end{Results*}
This is already the harshest scaling of dimension vs Schmidt number one can expect. That is, there exist states of arbitrarily high Schmidt number that admit a \emph{$4$-HQ} model, ruling out the possibility that Schmidt number can, in general, be device-independently certified in the most drastic sense. The Schmidt number of the states is unbounded, whereas device-independent quantification of a number above four is impossible. 

The argument extends to arbitrary sequential measurements, as transposed global sequential measurements are also valid measurements. More precisely, we have 
\begin{align}
    p(\vec{a} \vec{b}| \vec{x} \vec{y})=\text{Tr}( M_{\vec{a}|\vec{x}} \otimes  F_{b_1|y_1}^{(1)^{\dagger}} ...  F_{b_m|y_m}^{(m)^{\dagger}}  F_{b_m|y_m}^{(m)} ...  F_{b_1|y_1}^{(1)} \, \rho) \nonumber \\ 
    = \text{Tr}( M_{\vec{a}|\vec{x}} \otimes (F_{b_1|y_1}^{(1)^{\dagger}} ...  F_{b_m|y_m}^{(m)^{\dagger}}  F_{b_m|y_m}^{(m)} ...  F_{b_1|y_1}^{(1)})^T \, \rho^{T_B}) \nonumber \\ 
    = \text{Tr}( M_{\vec{a}|\vec{x}} \otimes (F_{b_1|y_1}^{(1)^{\dagger}})^* ...  (F_{b_m|y_m}^{(m)^{\dagger}})^*  (F_{b_m|y_m}^{(m)})^* ...  (F_{b_1|y_1}^{(1)})^* \, \rho^{T_B})
\end{align}
where $\{ M_{\vec{a}|\vec{x}} \}$ are Alice's global sequential measurements operators, $\{F_{b_k|m_k}^{(k)}\}$ are Bob's Kraus operators. Since for any set of Kraus operators $\{ F_l \}$ the complex conjugate $\{ F_l^* \}$ defines a valid set of Kraus operators (i.e. the normalisation is preserved), the proof \eqref{proof} extends to the sequential case. 

Moreover, since the Schmidt number is sub-multiplicative, it also means that an arbitrary number of copies $n$ of the state $\rho(d)$, always has a \emph{$4^n$-HQ} model, i.e. at most $4$ per copy, using the distributivity and tensor stability of partial transposition: 

\begin{align}
   ( \rho \otimes \rho ... \otimes \rho )^{T_B} = \rho^{T_B} \otimes \rho^{T_B}  ... \otimes \rho^{T_B} 
\end{align}
ruling out even the possibility of certifying the Schmidt number by coherently processing multiple copies at once.

It is curious that the partial transpose (and possibly other maps) can drastically change the entanglement content of a state, but it is not possible for them to induce entanglement. So, while our method of passing the dual of the positive map to the measurements can be used to rule out the possibility of device-independently quantifying entanglement, they cannot be adapted to answer the question of whether the presence of entanglement can always be device-independently certified. 

In summary, we have presented a proof that entanglement cannot be device-independently quantified, even when processing many copies simultaneously and for sequential measurements. While in the restricted Bell scenario, of single copy and non-sequential measurements, this already follows from \cite{Barrett02}, this is the first proof in a scenario where the general question of detection is still open.

Indeed, as we showcase in Fig. \ref{fig:Schematic}, processing multiple copies with potentially sequential measurements, is the most general scenario in which resources can be estimated without making further assumptions on the state or measurements. To violate a Bell inequality, multiple measurement rounds are usually employed to estimate joint probability distributions $p(ab|xy)$. So to ask, whether the entanglement of a quantum state $\rho$ can be device-independently characterised, can reasonably be interpreted as asking whether a suitably large number of states $\rho$ can violate a Bell inequality. Processing multiple copies then just implies additional experimental capabilities, but no further assumptions on the state of the system or any of the devices used.

There are two noteworthy comments to this paradigm. The first being finite statistics. If one only has a finite number of copies $n$ available, then accessing multiple copies at once could potentially adversely affect the statistical significance of certification. The other being an inherent assumption about an ability to distribute state $\rho$ identically and independently in each round. It may well be, that the physical state one determines in many rounds through tomography, is really just an average of very different state being produced in each round. In that case, having access to $\rho$, may not be identical to having access to $\rho^{\otimes k}$.  Nonetheless, if we ask: \emph{can the entanglement of all states $\rho$ be device-independently certified?}, the most natural interpretation of the question is the one we answer negatively in this paper.

While we believe that our result clarifies an important point on the relationship between non-locality and entanglement, the main question of device-independently certifying the presence of entanglement is still open. Additionally, in the context of quantification, it would be very interesting to know whether other measures of entanglement, such as e.g. the entanglement of formation or squashed entanglement, can also be altered by partial transposition and thus not be device-independently certified.

\emph{Acknowledgements:} We want to thank Jessica Bavaresco for inspiring presentations, Mateus Araújo and Nicolai Friis for suggesting promising future research directions and comments on an initial draft, and Fabien Clivaz for helpful feedback. We also acknowledge funding from FWF Y879-N27 (START) and the Swiss National Fund (SNF) through the early Postdoc.Mobility fellowship P2GEP2$\_$181509. 

\bibliography{bibliography}

\end{document}